\RequirePackage[2020-02-02]{latexrelease} 
\documentclass[aps,prl, twocolumn,longbibliography,10pt]{revtex4}

\usepackage{graphicx}
\usepackage{epstopdf}
\usepackage{amsmath}
\usepackage[colorlinks=true,citecolor=blue,urlcolor=magenta]{hyperref}
\usepackage{tabularx}

\begin{document}

\title{Vapor cell Rydberg atom electrometry with time-separated fields}

\author{Michael V. Romalis}
\email{romalis@princeton.edu} 
\affiliation{Department of Physics, Princeton University}

\author{Joe Wiedemann}
\affiliation{Department of Physics, Princeton University}

\author{Shaobo Zhang }
\affiliation{Department of Physics, Princeton University}

\author{Nezih Dural}
\affiliation{Department of Physics, Princeton University}

\date{\today}

\begin{abstract}
	
Rydberg atoms have large transition electric dipole moments and high sensitivity to electric fields. We describe a new method for microwave field sensing in a vapor cell consisting of separate excitation, quantum evolution between two Rydberg levels in the dark and state-dependent detection with probe laser transmission. Using microwave pulse techniques we study homogeneous and inhomogeneous relaxation of the coherence between Rydberg levels and demonstrate detection of a 10 GHz signal with electric field sensitivity of 10~nV/cm/Hz$^{1/2}$. 
\end{abstract}

\maketitle

 Atoms in highly-excited Rydberg states have large transition electric dipole moments and high sensitivity to electromagnetic waves with a frequency that matches energy separation between two nearby Rydberg levels. The possibility of building a practical electric field sensor using thermal alkali metal vapor in a glass cell has attracted significant attention \cite{Fan2015,RydESense,Rydmicrowave, Yuan2023}. Such experiments usually rely on electromagnetically-induced transparency (EIT) signals with two overlapping counter-propagating laser beams to excite and probe atoms in one of the two Rydberg states \cite{EITRyd}. The presence of a resonant electromagnetic wave generates Autler-Townes (AT) splitting of the Rydberg state that modifies the EIT spectrum \cite{EITAT}. Many modifications of this scheme have been explored to increase the sensitivity, including, for example, Mach–Zehnder interferometer for probe laser \cite{Homodyne}, three-laser interrogation \cite{Threeph,Threeph1}, introduction of a local auxiliary microwave field \cite{superhetero}, and sensing near critical point of many-body Rydberg interactions \cite{Critpoint}.

Here we describe a new method of Rydberg electrometry that separates the excitation, sensing and detection steps of the process. It allows one to measure coherent evolution between two Rydberg states in the dark, free from Doppler and laser linewidth broadening. Using pulsed techniques common in nuclear magnetic resonance (NMR) we can distinguish between homogeneous and inhomogeneous broadening of the Rydberg coherence and identify effects responsible for each. With the coherence time limited primarily by Rydberg-Rydberg collisions, we optimize sensitivity to weak electric fields by applying an excitation field that generates a large microwave coherence and realize sensitivity of 10 nV/cm/Hz$^{1/2}$, a significant improvement over current state of the art in vapor cell Rydberg electrometry \cite{superhetero,Repump,ThreeSens,Critpoint}.

Our approach only requires an addition of fast intensity control for the probe and coupling beams used in a typical 2-laser EIT setup. The probe laser excites alkali-metal atoms to the $P$ state, while the coupling laser excites them to one of the Rydberg states. Unlike EIT detection schemes, we do not rely on coherent evolution of the atoms in the two optical fields, but use them to simply excite atoms to the Rydberg state with a $ \sim 1~\mu$sec pulse of both lasers. Then we turn off both lasers and apply microwave pulses resonant with the energy separation between two Rydberg levels. In a simplified picture one can treat the two levels as a spin-1/2 system and apply pulses to manipulate their quantum state on the Bloch sphere. 

At the end of the microwave pulse sequence we measure the population in the first Rydberg state. We apply a $\sim 1~\mu$sec recycling pulse of the coupling laser to de-excite atoms from the Rydberg state to the $P$ state. They quickly decay to the ground state with a $\sim 30~$nsec radiative lifetime. We then turn on the probe laser for $\sim 1~\mu$sec to measure transmission through the alkali-metal vapor. If the  atoms were in the first Rydberg state at the end of the microwave pulse sequence, we expect to see reduced transmission, since the atoms are brought back to the ground state. If the atoms were in the second Rydberg state, they were not resonant with the recycling pulse of the coupling laser and remain in the Rydberg state, resulting in a greater transmission of the probe laser.

After a time comparable to atom transit time across the laser beams we can apply another pulse sequence. We typically compare probe transmission between a pair of pulse sequences with a difference in only one microwave pulse parameter, so changes in the probe transmission are associated only with Rydberg state evolution in the absence of light fields. 

Such Rydberg atom dark evolution sensing (RADES) approach can be used in thermal atomic vapors for a wide range of measurements.  We demonstrate a 3 $\mu$sec long Rabi precession in a microwave field, which allows one to extend frequency-based calibration of the microwave field strength to a sub-MHz regime without using a three-laser approach. We perform separate measurements of population decay time $T_1$, inhomogeneous relaxation time $T_2^*$ and homogeneous relaxation time $T_2$ of Cs Rydberg states and identify factors that affect each relaxation process. In contrast, typical Rydberg EIT experiments measure a single linewidth that is affected by all relaxation processes simultaneously as well as the laser linewidth and residual Doppler broadening. We also demonstrate sensing of weak microwave fields using this approach. To obtain first-order sensitivity of Rydberg population to a weak microwave field we apply a separate excitation microwave field that generates approximately $\pi/2$ tip on the two Rydberg state Bloch sphere. The strength of the excitation field is approximately 30 times smaller than in previous heterodyne sensing experiments \cite{superhetero,Repump}, indicating a large increase in the effective coherence time. 

\begin{figure}[!t]
	\centering
	\includegraphics[width=3in]{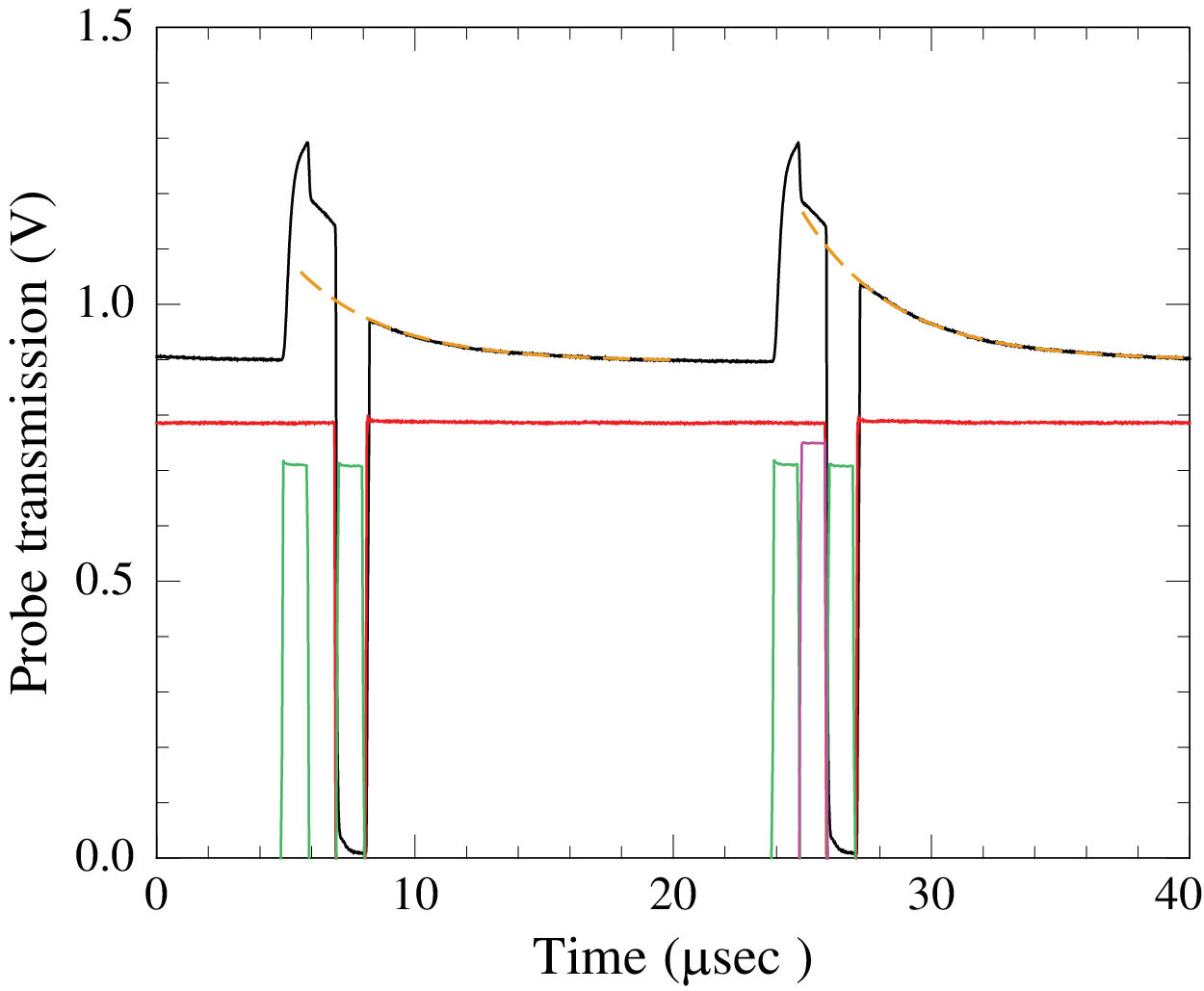}
	\caption{Sequence of pulses for Rydberg atom dark evolution sensing (RADES). Atoms are excited to Rydberg state by simultaneous red  and green laser pulse (green and red lines). Then green laser is turned off and a microwave pulse can be applied (purple line in second pulse sequence). The atoms that did not make a transition to another Rydberg state are recycled back to ground state with green pulse only and measured by monitoring probe transmission (black line) with the red laser turned on again. Dashed lines show extrapolation of the probe transmission transients, indicating the effect of the recycling pulse.
	}
	\label{Fig:MW recycle}
\end{figure}
{\it \noindent Experimental Setup.} The setup for our experiment is similar to common 2-beam vapor cell Rydberg electrometry setups with an addition of AOMs to control the intensity of the laser beams. The probe light at 852 nm is generated by a Moglabs ECDL, while the coupling light at 510 nm is produced by a Precilaser fiber amplifier and doubler seeded by Moglabs ECDL at 1020 nm. For some measurements the probe laser is locked to a Cs saturated absorption spectroscopy cell and the coupling laser is locked with an auxiliary EIT setup. However for many measurements we do not need to lock the lasers and simply scan one of them across the EIT peak since Rydberg coherent evolution is not affected by laser frequencies. We use fast AOMs operating near 200 MHz with tightly focused laser beams to obtain laser pulse rise time of 10-20 nsec. After AOMs the laser beams are collimated to a 2 mm $1/e^2$ diameter and counter-propagate through a 10 cm long 1" diameter cell. The incident power of the probe laser is 90 $\mu$W, the transmitted power is 20 $\mu$W. The coupling laser power is about 300 mW. The glass cell is fabricated by anodically bonding two windows with a double-sided, double-wavelength anti-reflection coating. The cell stem with Cs atoms is cooled to 10$^\circ$C by a thermoelectric cooler and the Cs density is measured to be $5\times 10^9/{\rm cm}^3$ by an absorption scan of the $6P_{3/2}$ $F=4$ resonance line. The magnetic field over the cell is controlled by a small set of 3-axis Helmholtz coils. The transmission of the probe laser is monitored by one channel of Thorlabs PDB450A photodetector. To compare probe transmission between two consecutive pulse sequences we use a Princeton Applied Research Boxcar Averager. It takes a difference between two transmission integration windows with fast re-trigger rates. Most of the measurements are performed on $42D_{5/2}\rightarrow 43P_{3/2}$ transition in Cs at 9.93 GHz.

{\it \noindent Measurements of recycling efficiency and population decay.} 
To determine the recycling efficiency of the coupling laser we measure the  transmission signal after the 2-laser excitation with and without the recycling pulse, as shown in Fig. 1. We find the recycling efficiency for our typical conditions is about 50\%. The recycling efficiency is sensitive to the magnetic field and  is maximized  when all components of the magnetic field are near zero. By delaying the  coupling laser recycling pulse we can measure population decay of the first Rydberg state. We find that $T_1$ relaxation time is about $4~\mu$sec, similar to the atom transit time across the laser beams.  

\begin{figure}[!t]
	\centering
	\includegraphics[width=2.5in,angle=90]{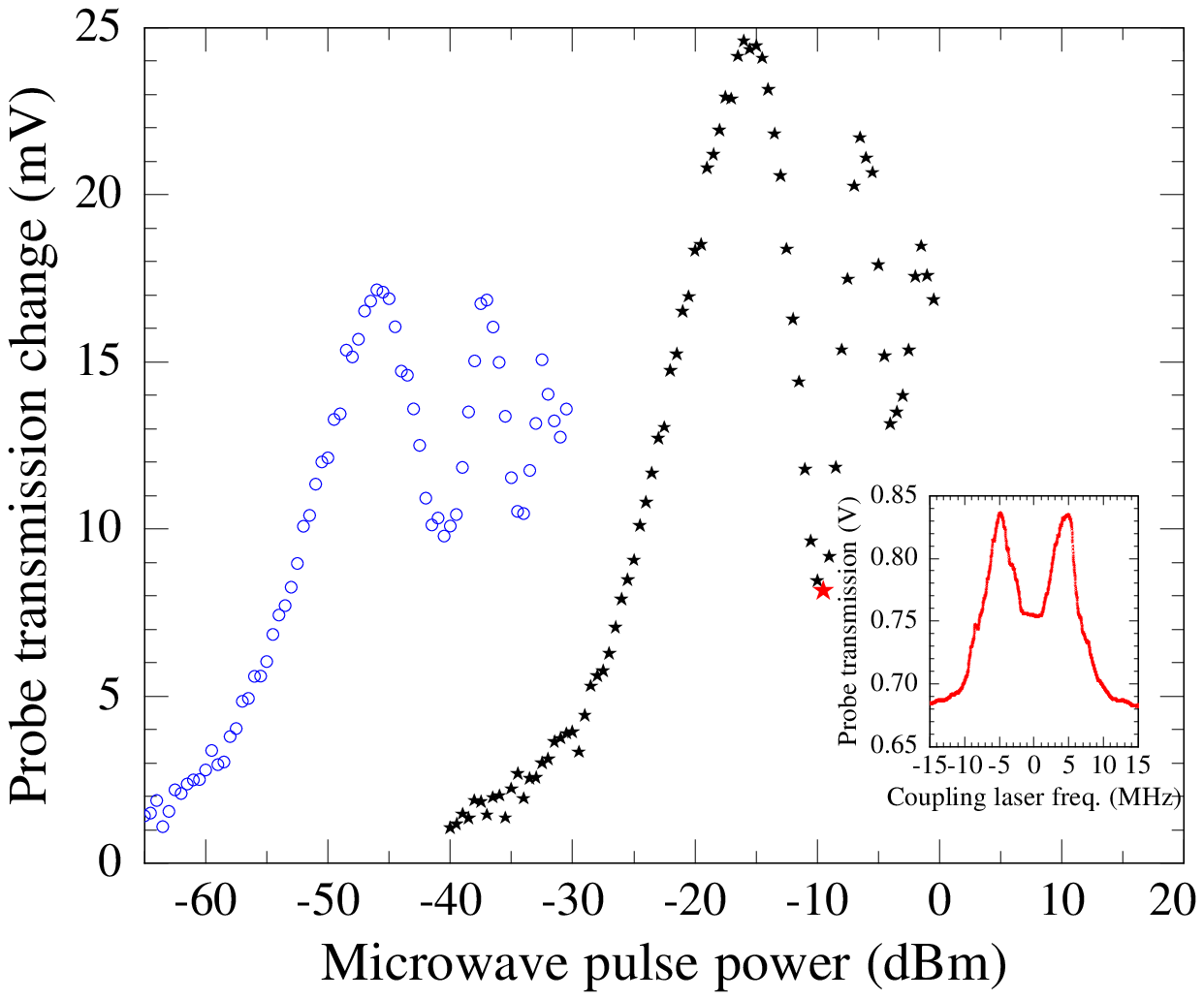}
	\caption{Measurements of coherent Rabi oscillations of Rydberg atoms in the dark. A microwave pulse of 100 nsec duration (filled black stars) or 3000 nsec (open blue circles) is applied with varying microwave power. An inset shows traditional EIT spectrum for -10 dBm microwave power with Autler-Townes splitting of 10 MHz. It corresponds to a $2 \pi$ rotation for an 100 nsec long pulse (red star).
	}
	\label{Fig:MWRabipulse}
\end{figure}

{\it \noindent Rabi precession measurements. }
Coherent Rydberg atom evolution in the presence of microwave fields can be observed by varying the amplitude or duration of the microwave pulse applied before the recycling pulse. An example of such measurement is shown in Fig. 2. We measure a change in probe transmission after the recycling pulse with and without the microwave pulse as the microwave power is varied for two pulse durations, 100 nsec and 3000 nsec. For 100 nsec long pulse the decrease in the Rabi oscillation contrast for larger rotation angles is primarily due to non-uniformity of the microwave field. For 3000 nsec microwave pulse one can see additional decay in the contrast due to  decoherence. For reference we also show a traditional EIT spectrum with 10 MHz A-T splitting that is obtained at the same microwave strength as a $2 \pi$ pulse with 100 nsec duration.  For longer pulses the pattern of Rabi oscillation is shifted to lower microwave power, in this case approximately 30 dB reduction. By measuring the period of Rabi oscillations one can extend precision determination of the microwave field strength to weaker fields when the Autler-Townes  splitting of the 2-laser EIT spectrum is no longer observable. For example, the first peak ($\pi$ rotation) for the 3000 nsec pulse signal corresponds to 166 kHz A-T splitting.

\begin{figure}[!t]
	\centering
	\includegraphics[width=5in,angle=90]{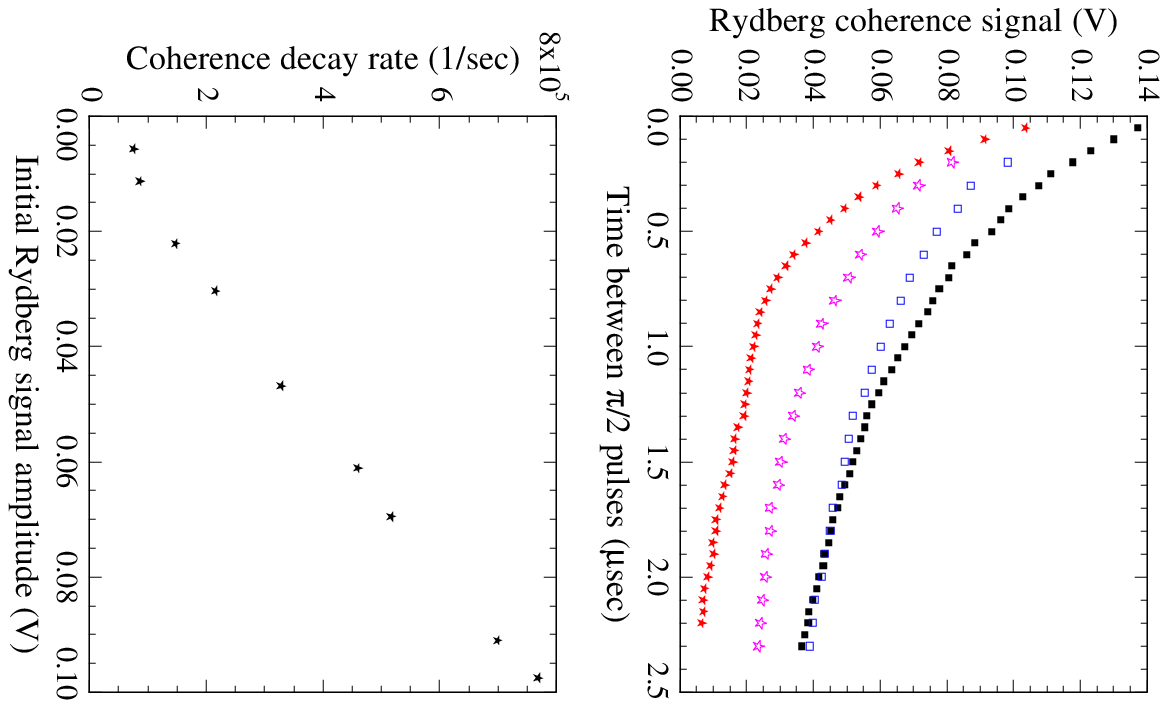}
	\caption{Top panel: Decay of coherence signal for two $\pi/2$ pulses (black filled squares) and  two $\pi/2$ pulses with a spin echo $\pi$ pulse in between  (blue open squares). Illumination of the cell with a red laser pointer causes much faster dephasing for two $\pi/2$ pulses (red filled stars), which can be partially recovered using spin-echo $\pi$ pulse (purple open stars).  Bottom panel: Coherence decay rate $T_2^{-1}$ as a function of initial Rydberg population proportional to the initial signal amplitude.
	}
	\label{Fig:T2meas}
\end{figure}

{\it \noindent Measurements of Rydberg coherence relaxation}. 
 We distinguish between inhomogeneous relaxation or dephasing time $T_2^*$ which corresponds to the loss of phase coherence across the whole cell and the homogeneous relaxation time $T_2$, which corresponds to random phase diffusion of individual atoms. For  $T_2^*$ measurements we use two $\pi/2$ pulses microwave pulses separated by a variable time $t$, for  $T_2$ measurements we add a spin-echo $\pi$ pulse at $t/2$. Microwave power is adjusted so $\pi/2$ pulse is about 10~nsec long and $\pi$ pulse is 20~nsec long. In order to reduce the effects of microwave field inhomogeneity, we also implement control of the phase of the microwave pulse using a fast voltage-controlled phase shifter and compare  pairs of microwave pulse sequences that differ only in the phase of the microwave field.  For  $T_2^*$ measurements we compare $\pi/2_x-t-\pi/2_x$ pulses with $\pi/2_x-t-\pi/2_{-x}$ pulses using a  $\pi$ phase shift for the second $\pi/2$ pulse. For $T_2$ measurement  we compare $\pi/2_x-t/2-\pi_y-t/2-\pi/2_x$ sequence with  $\pi/2_x-t/2-\pi_{-x}-t/2-\pi/2_x$ sequence by switching the phase shift of the $\pi$ pulse between $\pi/2$ and $\pi$.  The results of coherence measurements are summarized in Fig. 3. One can see that spin-echo coherence decay is slightly slower  than $T_2^*$ dephasing measurements. We find that ambient light illumination of the cell can have a significant effect on the dephasing time. To illustrate this point,  we illuminate a part of the cell with a non-resonant 650 nm laser pointer. This causes a dramatic reduction of the dephasing time, which can be partially recovered by using spin-echo $\pi$ pulse. This indicates that light illumination causes an inhomogenous dephasing, likely due to local electric fields. Electric fields can be generated in the cell due to photoelectric effect \cite{Photoelectric,Workfunction}.  We found that turning room lights off and minimizing laser scattering near the cell maximizes the dephasing time.
 
 To identify the source of Rydberg atom homogeneous decoherence  we vary the duration of the initial excitation pulse, using a shorter pulse reduces the initial density of Rydberg atoms and the size of the initial signal. Plotting the decoherence rate $1/T_2$ as a function of initial Rydberg signal amplitude reveals a linear relationship, indicating that the decoherence time is limited by Rydberg-Rydberg collisions. 
 
 \begin{figure}[!t]
 	\centering
 	\includegraphics[width=3in]{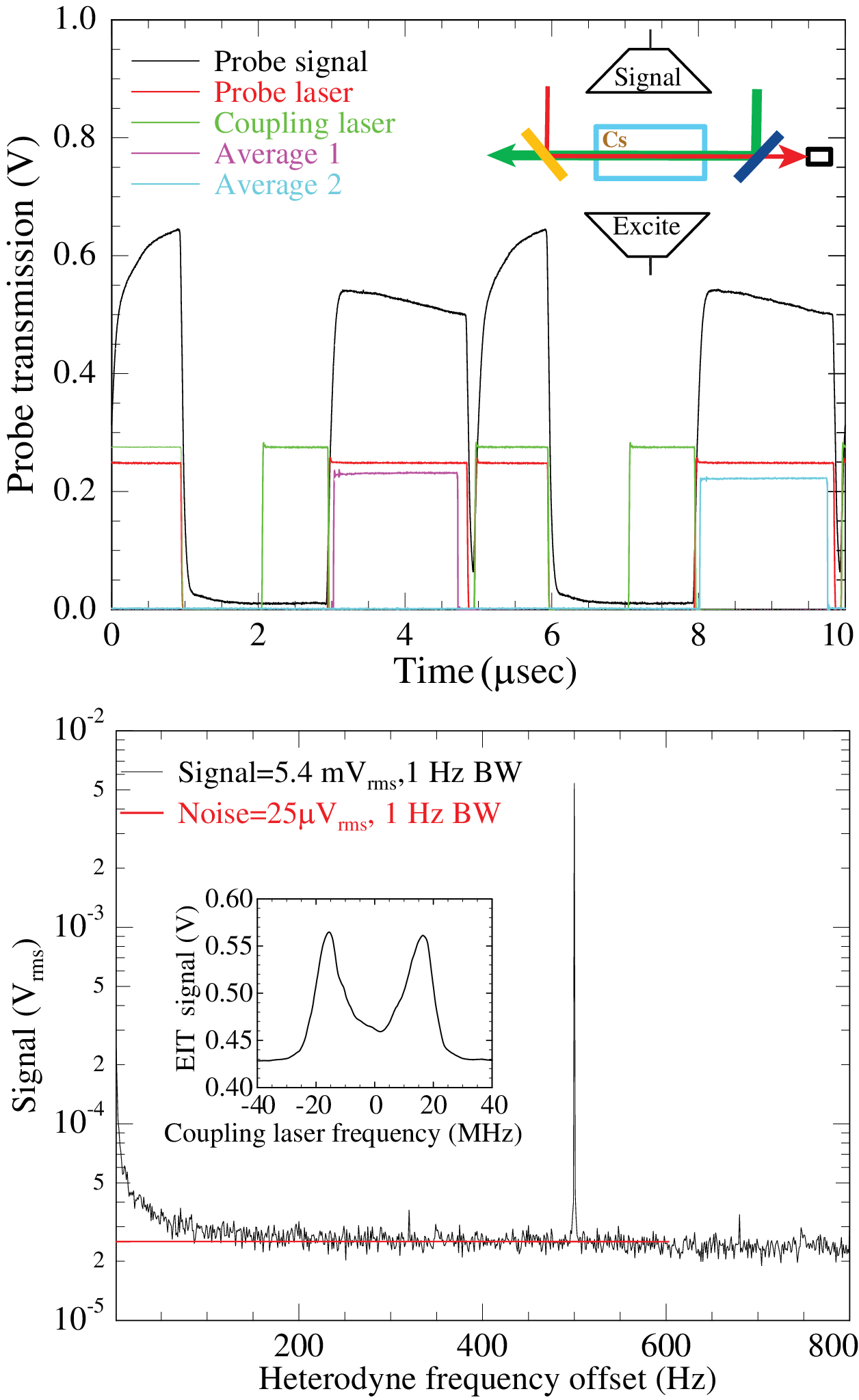}
 	\caption{Top Panel: Laser pulse sequence for weak microwave field sensing. The difference between Average 1 and Average 2, averaged over many pulses, is analyzed by the spectrum analyzer. Microwave fields are applied continuously, although most of the signal sensitivity comes from evolution during the dark periods. Inset shows arrangement of two microwave horns. Bottom Panel: Spectrum analyzer measurements for 1 Hz bandwidth and uniform window for a 2.2 $\mu$V/cm signal and 100 $\mu$V/cm excitation field. The inset shows calibration of the signal microwave horn giving a 32 MHz Autler-Townes splitting for an 80 dB greater microwave power. }
 	\label{Fig:T2meas}
 \end{figure}

{\it Measurements of weak microwave fields.}
For weak microwave sensing we implement a fast laser pulse sequence illustrated in Fig. 4 with a repetition rate of 5 $\mu$sec. Microwave fields are applied with two separate horns placed on opposite sides of the cell, as shown in the inset. One field serves to excite a finite microwave coherence, while the other provides a weak signal to be detected. The frequencies of the two microwave fields are detuned by 100500 Hz. The detuning of 100 kHz generates an opposite phase between the excitation and the signal microwave fields for each pulse sequence, so that there is alternatively greater and smaller tipping angle for the microwave coherence. An additional detuning of 500 Hz provides a modulation signal that can be detected with a spectrum analyzer. The excitation microwave field is applied to tip the spins by approximately $\pi/2$, which maximizes Rydberg population sensitivity to an additional small tipping angle created by the signal microwave field. We adjust the excitation field amplitude to maximize the signal and find that the optimal field is equal to 100 $\mu$V/cm. It corresponds to 145 kHz Rabi frequency, which gives 0.91 rad tipping angle in 1 $\mu$sec. The effective evolution time for the microwave coherence is somewhat longer than 1~$\mu$sec due to the finite duration of the excitation and recycling pulses, so the tipping angle is closer to 1.5 rad as expected for optimal sensitivity.  The strength of the signal microwave field is calibrated using Autler-Townes splitting by scanning the coupling laser frequency. For the  $42D_{5/2}\rightarrow 43P_{3/2}$ transition in Cs at 9.93 GHz we use dipole moment corresponding to a splitting of 14.6 MHz in a 1 V/m electric field \cite{DualRyd,Arc}. For signal electric field amplitude equal to 2.2 $\mu$V/cm we observed SNR=216 in the frequency spectrum with 1 Hz resolution, as shown in Fig. 4, corresponding to sensitivity of 10 nV${_p}$/cm/Hz$^{1/2}$ or 7 nV$_{rms}$/cm/Hz$^{1/2}$. This can be compared to the best previous heterodyne measurements with a measured noise spectral density \cite{superhetero}, equal to 55 nV$_{rms}$/cm/Hz$^{1/2}$. In our experiment the optimal excitation electric field amplitude is 30 times smaller (100 $\mu$V/cm vs. 3 mV/cm), indicating a roughly 30 times reduction in the effective linewidth. We also use much simpler laser frequency locking scheme and no additional laser intensity stabilization.

{\it Summary.} We have presented a new method for Rydberg atom interrogation in a thermal vapor  based on sequential application of laser fields and microwave fields. A crucial aspect of this method is state-selective de-excitation of the Rydberg atoms using the coupling laser. This allows one to transfer the information about Rydberg state population to the ground state, where it can be measured using transmission of the probe laser. Measurements of the microwave coherence in the dark allows one to distinguish between different relaxation processes,  described  by $T_1$, $T_2^*$ and $T_2$ time constants. We find that $T_1$ is dominated by beam transit time for our conditions, $T_2^*$ is affected by ambient light illumination of the cell, while  $T_2$ is limited by Rydberg-Rydberg collisions. To detect weak microwave signals we implement a fast pulse sequence that can detect small differences in the tipping angle of the microwave coherence.

Since the microwave coherence time is limited by collisions between Rydberg atoms, further improvements in sensitivity can be obtained by increasing the active volume of the measurement region, which was equal to 0.3~cm$^3$ in our setup, as well as increasing the probe transmission signal contrast. While we tested detection of continuous microwave fields, pulsed fields with $\sim 1~ \mu$sec duration can be detected without loss of sensitivity if timed to correspond to dark sensing intervals.  One can also adjust the sensor bandwidth by operating at higher or lower Rydberg density with corresponding shorter or longer pulse intervals.  
 
This work was supported by DARPA SAVANT program. We acknowledge helpfull discussions with James Shaffer and Kaitlin Moore.

\bibliography{RydbergReferences.bib}

\end{document}